\documentclass[pre,twocolumn,showpacs,showkeys,preprintnumbers,amsmath,amssymb]{revtex4-1}
\usepackage{graphicx}
\usepackage{bm}
\begin{document}

\title{Effective-field theory study of the dynamical Ising-type thin films}

\author{Bahad\i r~Ozan~Akta\c{s}}
\affiliation{Dokuz Eyl\"{u}l University, Graduate School of Natural and Applied Sciences, TR-35160 Izmir, Turkey}
\author{{\" U}mit Ak\i nc\i}
\author{Hamza Polat}
\email{hamza.polat@deu.edu.tr}
\affiliation{Department of Physics, Dokuz Eyl\"{u}l University, TR-35160 Izmir, Turkey}

\date{\today}

\begin{abstract}
The stationary state solutions of the Ising-type thin films with different layers in the presence of an external oscillatory field are examined within the effective-field theory. The study focuses on understanding of the effects of external field frequency and amplitude on the overall behavior. The particular attention is paid on evolution of the special point with dynamic field frequency corresponding the critical temperature of the three-dimensional infinite bulk system where the surface and modified exchange parameters are of no importance. Some outstanding findings such as surface enhancement phenomenon, effect of thickness on the dynamic process etc. are introduced together with well known other characteristics. An attempt is made to explain the relations between the competing time scales (intrinsic microscopic relaxation time of the system and the time period of the external oscillatory field) and frequency dispersion of the critical temperature coordinate of the special point.
\end{abstract}


\maketitle

\section{Introduction}\label{intro}
The conceptual frame of surface enhancement phenomenon on the finite magnetic materials, especially on semi-finite systems and thin films, has not lost its currency yet and recently there has been growing interest for this mature subject of solid state physics because a variety of apparently very different phenomena can be described by both experimentalists and theorists \cite{pleimling}. Thin film of nanometer thickness has been fabricated by various coating techniques and has many applications in technology. Advances in experimental techniques today allow to crosschecking of such far-reaching predictions. In the past quarter century, some preliminary experimental evidence has been presented for the existence of a surface type of magnetic ordering since the magnetic properties of free surfaces drastically differ from the bulk material, because the free surface breaks the translational symmetry, i.e. surface atoms are embedded in an environment of lower symmetry than that of the inner atoms and consequently the exchange constants between atoms in the surface region may differ from the bulk value.

One of the early experimental studies dating back to the last quarter of the century, has been made by Weller et al. \cite{weller}. There exists an empirical inference that the 4$f$ spins of the surface are not ferromagnetically coupled to the bulk moments. Due to this finding, the relevant work involves ground-breaking discoveries about the understanding of surface-enhanced magnetic order and magnetic surface reconstruction on Gd(0001). They have presented spin-polarized low-energy-electron diffraction (SPLEED) and magneto-optical Kerr effect (MOKE) studies performed \emph{in situ} that demonstrate the existence of surface-enhanced magnetic order. Their results had made a new Monte Carlo (MC) calculation for systems with antiferromagnetic (AF) perpendicular surface layer coupling compulsory. Also, D{\" u}rr et al. \cite{durr} have showed in the same period that the temperature dependence of the long-range order parameter in thin (1-3 ML) films of bcc Fe on Au(100)$-$including one monolayer$-$, as measured with SPLEED and spin-polarized secondary electron emission spectroscopy (SPSEE) as a complement. Their work represented a physical realization of a truly two-dimensional phase transition. Additionally they reported that the thickness independence of the critical behavior confirmed the universality hypothesis. Two unassailable experimental observations published in the early of two decades ago. One of them includes an \emph{in situ} magnetic resonance (MR) measurements of ultrathin Ni(110) which was prepared in ultrahigh vacuum. Li and Baberschke \cite{li1} have carried out a comprehensive investigation to determine the thickness dependence of the critical exponent $\beta$ and the Curie temperature $T_{c}$. They concluded that the existence of a crossover in the magnetic properties of Ni films as a function of thickness and also the critical exponents have been determined in their experiment agree very well with the known exponents for a corresponding 2D Ising-type bulk Ni. In the other one, thickness-dependent Curie temperature of 5-100 monolayers Gd on W(100) and its dependence on the growth conditions are determined by \emph{in situ} ac-susceptibility ($\chi_{ac}$) measurements by Farle et al. \cite{farle1}. They found that the Curie temperature of carefully prepared layer-by-layer-grown Gd(0001) films varies from $T_{c}(\mathrm{bulk})=292.5$K to $T_{c}(5\mathrm{ML})=120$K and it was also demonstrated that different growth conditions change the film's Curie temperature dramatically. After the aforementioned outstanding studies, a comprehensive examination has been propounded in the subsequent years by Poulopoulos and Baberschke \cite{poulopoulos}. They discussed the fundamental magnetic observables, i.e. magnetic moment per atom, Curie temperature, susceptibility and magnetic anisotropy, for idealized prototype thin films like Fe, Co, Ni on metal substrates such as Cu, W, Re. They also presented in depth satisfactory knowledge about studies on trilayers, i.e. magnetic thin films separated by a spacer, like Cu. Apart from these, the interesting phenomenon of surface enhanced magnetic order, i.e. the coexistence of an ordered surface with a disordered bulk has been observed in two of the 4$f$ rare-earth metals: Gd and Tb. Former experimental measurements suggest the possibility of large surface anisotropies in thin films \cite{jonker, pescia1, pierce, pappas}. Beside the relevant elements, magnetically ordered surface can exist also for Cr \cite{rau1, rau2}. It is possible to say that there exists a general phenomenon due to the reduced coordination number, the critical temperature is lower at surface in thin films and decreases with decreasing film thickness \cite{stampanoni1, przybylski, schneider}. As presented by Detzel et al. \cite{detzel} and Stampanoni et al. \cite{stampanoni2}, for fcc Fe films on Cu(100) substrates, the Curie temperature increases strongly from one to two monolayers, but decreases in thicker layers. For the Fe films grown on Ni/Si substrates, it has been observed that a transition between magnetic and nonmagnetic phases occurs at a critical thickness \cite{edelstein, krishnan, colombo, li2}. With the recent advances in epitaxial growth techniques (especially in molecular beam epitaxy) it is now possible to grow very thin magnetic films of controllable thickness and this has stimulated renewed interest in both experimental and theoretical film magnetism \cite{farle2, pescia2}.

Until recently, theoretical treatments on the semi-infinite systems and thin films in the literature have continued simultaneously with experiments \cite{binder1, binder2, aguilera-granja, landau, wang, yao, saber1, laoisiritaworn1, cossio, laoisiritaworn2, park}. Several theoretical and/or computational studies on pure crystalline thin films have been put forward to explain nature of surface enhancement mechanism since the early experimental observations. First, by Binder and Hohenberg \cite{binder1}, Ising models with modified exchange interaction on the surface have been considered. The critical value of the modified exchange for surface ordering has been found from high-temperature-series expansion (HTSE) and compared to mean-field theory (MFT) value in their pioneering work \cite{binder1}. They reported that there was a temperature region in which the surface behaves like a two-dimensional Ising-type bulk near its transition point for the values greater than the critical value of relevant exchange. According to Binder and Hohenberg, the critical exponents experience a crossover at the critical value of the exchange and effective exponents differ from the true ones for values less than the critical value of the exchange. The average magnetization of the simple cubic Ising films (with different thicknesses atomic layers) has been calculated as a function of temperature using the MC technique by Binder in the same year \cite{binder2}. The semi-infinite Ising model with an arbitrary number of surface magnetic couplings differing from the bulk exchange constant, has been solved within the MFT by Aguilera-Granja and Mor\'{a}n-L\'{o}pez in the middle of three decades ago \cite{aguilera-granja}. They showed that the exact expressions for the critical couplings would lead to a higher surface Curie temperature than the bulk one. Landau and Binder \cite{landau} have realized a MC simulations and they have presented the results of phase transitions and critical behavior at the surface of a simple cubic Ising model. The work includes detailed extraction of surface and bulk properties as a function of temperature and surface coupling. Also, the surface-bulk multicritical point is located with improved precision and crossover behavior has been studied.  Later on, a review of surface magnetism has been given by Kaneyoshi \cite{kaneyoshi4}. After a survey of experimental and theoretical results for magnetic moment and anisotropy at surfaces or interfaces of semi-infinite magnets and thin films in the article, the interplay of magnetizations and anisotropy at a surface has been discussed. Ferroelectric films have been studied to obtain the polarization and the dependence of the Curie temperature on the long-range exponent in two works \cite{wang, yao}. A cubic ferroelectric lattice was assumed to consist of pseudo-spins with interactions. Parallel study using a distorted lattices, such as tetragonal lattice, has been gave the same dependence of the Curie temperature and polarization according to Wang et al. \cite{wang}. Yao et al. \cite{yao} has presented that the long-range dipole-dipole interaction made the transition temperature higher. In both studies \cite{wang, yao}, films have been described by transverse Ising model (TIM) with long-range interactions. The phase transitions of in a transverse spin-1/2 Ising film has been also examined in detail within the framework of the effective-field theory (EFT) by Saber et al. \cite{saber1}. It has been found that if the ratio of the surface interaction to the bulk one is less than a critical value, the critical temperature of the film is smaller than the bulk critical temperature and as the film thickness is increased further, critical temperature increases and approaches asymptotically to the bulk-critical temperature for large values of the thickness. Additionally they reported that the critical temperature is larger than both bulk and surface critical temperatures of the corresponding semi-infinite system. As the film thickness increases further, critical temperature decreases and approaches asymptotically to the bulk one. Following this, for large values of the thickness, the phase transitions on surface observed for greater values of the critical value of modified exchange in the corresponding semi-infinite system  \cite{saber1}. Also, the layer magnetizations and their profiles have been presented as an essential detail and they have illustrated the existence of one defined critical temperature of the film. In MC simulations on semi-infinite systems, a linear dimension parameter should be much larger than the other two to maintain the quasi two dimensional structure of the films. This detail should also be taken in consideration to preserve the films' geometry. Within this computational framework, in recent years, Laosiritaworn and co-workers \cite{laoisiritaworn1} have used MC simulations and the MFT to observe the magnetic behavior of Ising films with cubic lattice structure as a function of temperature with thickness. They found that the magnetic behavior changes from two-dimensional to the three-dimensional character with increasing film thickness and this finding made great strides in semi-infinite concept. Both the crossover of the critical temperature from a two-dimensional to a bulk value and shift exponent have been observed. The same computational technique has been used by Cossio et al. \cite{cossio} to explain the temperature dependence of the magnetization, the magnetic susceptibility and also the fourth-order Binder's cumulant. Ferromagnetic (F) Ising-type thin film in the presence of a time dependent external oscillatory magnetic field has been investigated to model the hysteretic behavior of the system by Laosiritaworn \cite{laoisiritaworn2}. In the work, MC simulation has been used and thickness dependence of hysteresis properties for varying frequency and amplitude of the external field has been investigated by the author. The power law scaling relations among the hysteresis properties, the thickness and the field parameters have been found i.e. scaling of the hysteresis loop (HL) area performed and scaling exponents were also reported. In light of those results, it appears that hysteresis properties of thin films under external perturbation can be predicted which provides another successful step in modeling ferromagnetic materials. In a brand new paper, the first study of the surface critical properties at a dynamic phase transition has been presented by Park and Pleimling \cite{park}. Both in two and three space dimensions they have obtained values for the surface critical exponents that differ markedly from the values of the equilibrium surface exponents, thus demonstrating that the dynamic surface universality class differs from that of the equilibrium system, even though the same universality class prevails for the corresponding bulk systems. However, the rich behavior depending on the dynamical parameters has not been covered in both works \cite{laoisiritaworn2, park}. Very recently, the effect of the random magnetic fields distributed by a Gaussian distribution centered at zero on the phase diagrams and ground state magnetization of the thin film described by TIM has been investigated by Ak\i nc\i~\cite{akinci}. Particular attention has been paid on the evolution of the special point coordinate with distribution parameter in relevant study. The author found that rising the distribution parameter makes no significant change in special point in both planes, but it gives rise to a decline in coordinates and after a certain value of distribution parameter, the special point disappeare in both planes simultaneously.

In the semi-infinite systems, depending on the ratio between surface and bulk exchange interactions, the system may order on the surface before it orders in the bulk which is called extraordinary transition. In the contrary of this, ordinary transition means that the surface critical temperature is the same with the bulk transition temperature. There exists an exact consensus in literature that, the intersection point between these two transitions is called `special point' (in other words `crossover point') \cite{li1, binder1, landau, saber1, laoisiritaworn1, cossio, park, akinci, zaim, kaneyoshi3}. As the film gets thicker, it approaches the semi-infinite system, and this unusual effect shows itself as an intersection point in the phase diagrams plotted in critical temperature versus surface exchange interaction plane for the films with different thicknesses.

It is well known that the EFT is one of the most powerful methods that determines the boundary which separates several phases in the relevant planes, based on the use of rigorous correlation identities as a starting point and utilizes the differential operator technique firstly developed by Honmura and Kaneyoshi \cite{honmura}. Although the conventional version of the method fails to find an expression for the free energy, since it takes into account the self spin correlations, the method is superior to MFT which neglects the thermal fluctuations via neglecting the self spin correlations. Thus, it is expected from EFT to obtain more reasonable results than MFT for these systems, as in the case of static Ising model. Similar to the our problem, Shi and co-workers have carried out first EFT study of a bulk system with a kinetic Ising-type Hamiltonian \cite{shi}. Therefore in this work, we intend to probe the phenomena mentioned above of Ising-type semi-infinite systems and thin films in the presence of a time-dependent external oscillatory magnetic field by using the EFT. These types of perturbations constitute an important role in material science, since the time-dependent external effects may reveal the origin of some important macroscopic behavior pattern which is still open for inspection in intrinsic surface magnetism. Eventually, for these purposes the outline of the article is as follows: We briefly describe the formalism and the method used for the system in Sec. 2. Numerical results and discussions are summarized in Sec. 3, and finally Sec. 4 contains remarks about our conclusions.

\section{Methodology}\label{method}
\begin{figure}[b]
\includegraphics[width=8.5cm]{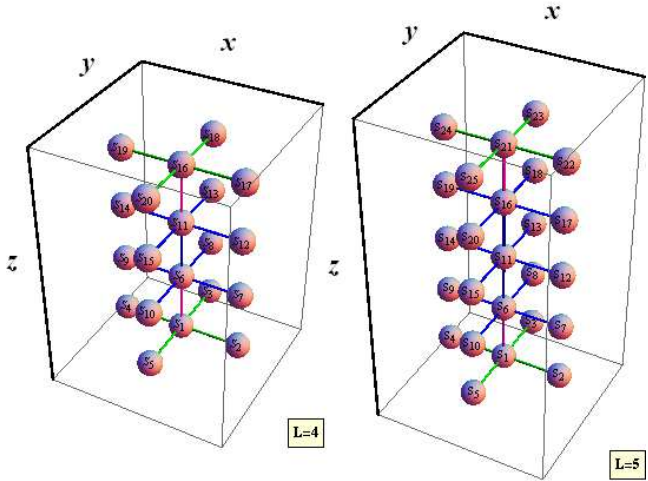}\\
\caption{(Color online) Magnetic unit cell for the simple cubic thin films of two different thicknesses as $L=4,$ $5$. The lattice has a finite size only in $z$ direction as other two directions are defined on the interval such that $x,y$ $\epsilon$ $(-\infty,\infty)$.  Green, magenta and blue cylindrical edge rendering functions represent $J_{1},$ $J_{2}$ and $J_{3}$ ferromagnetic exchanges respectively, as well as spherical vertices symbolize the magnetic items.}\label{fig1}
\end{figure}

In order to investigate the dynamical transitions, surface enhancement phenomenon, dynamical symmetry breaking, strength of the magnetic order in different layers and many other dynamical features, one simple choice may be a dynamical Ising-type simple cubic isometry which has an inner coordination number $z=4$ defined on the three dimension with a time dependent external oscillatory (in time but uniform over the space) magnetic field studied by EFT. For this purpose, we consider the following Hamiltonian,
\begin{equation}\label{eq1}
\mathcal{H}=-\sum_{\langle ij\rangle}J_{ij}s_{i}s_{j}-h(t)\sum_{i} s_{i}
\end{equation}
where $s_{i}$ is the spin operator at a lattice site $i$ and any spin variable can take the values $s_{i}=\pm 1$. As is known, $\langle \ldots\rangle$ subscript bracket symbolize the nearest neighboring in first summation. The second summation is over all the lattice sites. The exchange interaction $J_{ij}$ between the spins on the sites $i$ and $j$ takes the values according to the positions of the nearest neighbor spins. Two surfaces of the film have the intralayer coupling $J_{1}$. The interlayer coupling between the surface and its adjacent layer (i.e. layers $1,$ $2$ and $L,$ $L-1$) is denoted by $J_{2}$. For the rest of the layers, the interlayer and the intralayer couplings are assumed as $J_{3}$. The system has three exchange interactions where $J_{1},$ $J_{2},$ $J_{3}>0$ favors a ferromagnetic leaning of the adjacent sites as shown in Fig. (\ref{fig1}) and the Zeeman term describes interaction of the spins with the field of the sinusoidal form
\begin{equation}\label{eq2}
h(t)=h_{0}\cos(\omega t),
\end{equation}
where $t$ is the time and $h_{0}$ is the amplitude of the oscillatory magnetic field with an angular frequency $\omega$.

Conventionally, the layer agents remaining invariant under the appropriate symmetry operation in a bulk system are defined with certain criterions: Spin values, types and number of it's neighboring interactions. The relevant agents depict thermo-magnetical behavior of whole material whereas in a little more realistic systems (such as magnetic thin films), one should construct the equations of state for each layer agents to control the dependency of the characteristics with the film thickness hence, such criterion-related validity is lost. Our system is in contact with an isothermal heat bath at given temperature $T$. So, the dynamical evolution of the system may be given by non-equilibrium Glauber dynamics \cite{glauber} based on a master equation. The dynamical equations of motion for each layer are in the form of
\begin{equation}\label{eq3}
\tau \frac{d}{dt}\langle s_{i}^{(k)}\rangle=-\langle s_{i}^{(k)}\rangle+\langle\tanh[\beta(E_{i}^{(k)}+h(t))]\rangle,\hspace{0.2cm} k=1,\ldots,L
\end{equation}
with
$$$$
\begin{eqnarray}\label{eq4}
E_{i}^{(1)} \hspace{0.48cm}& = & \sum_{\delta=1}^{z}J_{1}s_{\delta}^{(1)}+J_{2}s_{\delta'}^{(2)}, \nonumber\\
E_{i}^{(2)} \hspace{0.48cm}& = & J_{2}s_{\delta'}^{(1)}+\sum_{\delta=1}^{z}J_{3}s_{\delta}^{(2)}+J_{3}s_{\delta'}^{(3)}, \nonumber\\
\quad\vdots \hspace{0.6cm}& \vdots & \quad\quad\quad\quad\quad\quad\quad\vdots \nonumber\\
E_{i}^{(k)} \hspace{0.35cm}& = & J_{3}s_{\delta'}^{(k-1)}+\sum_{\delta=1}^{z}J_{3}s_{\delta}^{(k)}+J_{3}s_{\delta'}^{(k+1)}, \nonumber\\
\quad\vdots \hspace{0.6cm}& \vdots & \quad\quad\quad\quad\quad\quad\quad\vdots \nonumber\\
E_{i}^{(L-1)} & = & J_{3}s_{\delta'}^{(L-2)}+\sum_{\delta=1}^{z}J_{3}s_{\delta}^{(L-1)}+J_{3}s_{\delta'}^{(L)}, \nonumber\\
E_{i}^{(L)} \hspace{0.38cm}& = & J_{2}s_{\delta'}^{(L-1)}+\sum_{\delta=1}^{z}J_{1}s_{\delta}^{(L)},
\end{eqnarray}
where $1/\tau$ is the transition per unit time in a Glauber type stochastic process, $\beta=1/k_{B}T$ and $k_{B}$ represents the Boltzmann constant. The subscript and superscript in a spin variable symbolizes the lattice site and layer number respectively as well as $E_{i}^{(k)}$ is the local field acting on the site $i$ in $k$th layer. $\delta$ and $\delta'$ subscripts in the energy contributions represent in-layer and out-of-layer nearest neighboring and this type of indication also is reinforced by superscripting with regard to control the relation between production index and the nearest neighboring argument in progressive aspects. The average brackets in Eq. (\ref{eq3}) stand for the usual canonical thermal average.

In order to handle the second term on the right-hand side of the Eq. (\ref{eq3}) one can use the differential operator technique \cite{honmura, kaneyoshi1}. By using the technique, Eq. (\ref{eq3}) gets in the form of
\begin{equation}\label{eq5}
\tau \frac{d}{dt}\langle s_{i}^{(k)}\rangle=-\langle s_{i}^{(k)}\rangle+\langle\exp(E_{i}^{(k)}\nabla)\rangle f(x)|_{x=0},\hspace{0.2cm} k=1,\ldots,L
\end{equation}
where $\nabla=\partial/\partial x$ is one dimensional differential operator and the function $f(x)$ is given by
\begin{equation}\label{eq6}
f(x)=\tanh[\beta(x+h(t))],
\end{equation}
as for that the effect of the differential operator on a function $f(x)$,
\begin{equation}\label{eq7}
\exp(a\nabla)f(x)|_{x=0}=f(x+a)|_{x=0},
\end{equation}
with any real constant $a$. By using the energy expression given in Eq. (\ref{eq4}), in Eq. (\ref{eq5}) then, in order to get a polynomial form of the second term on the right-hand side of the equations, by using the van der Waerden identity for two-state spin, i.e., $\exp(bs_{i})=\cosh(b)+s_{i}\sinh(b)$ where $b$ is any real constant in a representative manner, we write the exponential term in terms of the hyperbolic trigonometric functions, Eq. (\ref{eq5}) exactly written in terms of multiple spin correlation functions occurring on the right-hand side. Thus we get
\begin{widetext}
\begin{eqnarray}\label{eq8}
\tau \frac{d}{dt}\langle s_{i}^{(1)}\rangle&=&-\langle s_{i}^{(1)}\rangle+\left\langle \prod_{\delta=1}^{z}[A_{1}+s_{\delta}^{(1)}B_{1}][A_{2}+s_{\delta'}^{(2)}B_{2}]\right\rangle,\nonumber\\
\tau \frac{d}{dt}\langle s_{i}^{(2)}\rangle&=&-\langle s_{i}^{(2)}\rangle+\left\langle [A_{2}+s_{\delta'}^{(1)}B_{2}]\prod_{\delta=1}^{z}[A_{3}+s_{\delta}^{(2)}B_{3}][A_{3}+s_{\delta'}^{(3)}B_{3}]\right\rangle,\nonumber\\
\vdots\qquad&\vdots&\quad\quad\quad\quad\quad\quad\quad\quad\quad\quad\quad\quad\quad\quad\quad\quad\vdots\nonumber\\
\tau \frac{d}{dt}\langle s_{i}^{(k)}\rangle&=&-\langle s_{i}^{(k)}\rangle+\left\langle [A_{3}+s_{\delta'}^{(k-1)}B_{3}]\prod_{\delta=1}^{z}[A_{3}+s_{\delta}^{(k)}B_{3}][A_{3}+s_{\delta'}^{(k+1)}B_{3}]\right\rangle,\nonumber\\
\vdots\qquad&\vdots&\quad\quad\quad\quad\quad\quad\quad\quad\quad\quad\quad\quad\quad\quad\quad\quad\vdots\nonumber\\
\tau \frac{d}{dt}\langle s_{i}^{(L-1)}\rangle&=&-\langle s_{i}^{(L-1)}\rangle+\left\langle [A_{3}+s_{\delta'}^{(L-2)}B_{3}]\prod_{\delta=1}^{z}[A_{3}+s_{\delta}^{(L-1)}B_{3}][A_{2}+s_{\delta'}^{(L)}B_{2}]\right\rangle,\nonumber\\
\tau \frac{d}{dt}\langle s_{i}^{(L)}\rangle&=&-\langle s_{i}^{(L)}\rangle+\left\langle [A_{2}+s_{\delta'}^{(L-1)}B_{2}]\prod_{\delta=1}^{z}[A_{1}+s_{\delta}^{(L)}B_{1}]\right\rangle,
\end{eqnarray}
\end{widetext}
with
\begin{eqnarray}\label{eq9}
A_{n}&=&\cosh(J_{n}\nabla)f(x)|_{x=0},\nonumber\\
B_{n}&=&\sinh(J_{n}\nabla)f(x)|_{x=0},
\end{eqnarray}
where $n=1,2,3$.

When the product in Eq. (\ref{eq8}) is expanded, the multi site spin correlations appear. In order to make the expansion manageable let us handle these correlations with a decoupling approximation (DA) \cite{tucker} as
\begin{equation}\label{eq10}
\langle s_{i}^{(k)}\ldots s_{j}^{(k)}\ldots s_{l}^{(k)}\rangle=\langle s_{1}^{(k)}\rangle\ldots \langle s_{j}^{(k)}\rangle \ldots \langle s_{l}^{(k)}\rangle.
\end{equation}
In fact, the primitive form of this improvement corresponds essentially to the Zernike approximation \cite{zernike} in the bulk problem, and has been successfully applied to a great number of magnetic systems including the surface problems \cite{balcerzak, kaneyoshi2, kaneyoshi3}. By using the expansion in Eq. (\ref{eq10}) and the assumption below
\begin{equation}\label{eq11}
\langle s_{i}^{(k)}\rangle=m_{k},\hspace{0.2cm} k=1,\ldots,L,
\end{equation}
the Eq. (\ref{eq8}) morph into the form of,
\begin{widetext}
\begin{eqnarray}\label{eq12}
\tau \frac{dm_{1}}{dt}&=&-m_{1}+[A_{1}+m_{1}B_{1}]^{z}[A_{2}+m_{2}B_{2}],\nonumber\\
\tau \frac{dm_{2}}{dt}&=&-m_{2}+[A_{2}+m_{1}B_{2}][A_{3}+m_{2}B_{3}]^{z}[A_{3}+m_{3}B_{3}],\nonumber\\
\vdots\quad&\vdots&\quad\quad\quad\quad\quad\quad\quad\quad\vdots\nonumber\\
\tau \frac{dm_{k}}{dt}&=&-m_{k}+[A_{3}+m_{k-1}B_{3}][A_{3}+m_{k}B_{3}]^{z}[A_{3}+m_{k+1}B_{3}],\nonumber\\
\vdots\quad&\vdots&\quad\quad\quad\quad\quad\quad\quad\quad\vdots\nonumber\\
\tau \frac{dm_{L-1}}{dt}&=&-m_{L-1}+[A_{3}+m_{L-2}B_{3}][A_{3}+m_{L-1}B_{3}]^{z}[A_{2}+m_{L}B_{2}],\nonumber\\
\tau \frac{dm_{L}}{dt}&=&-m_{L}+[A_{2}+m_{L-1}B_{2}][A_{1}+m_{L}B_{1}]^{z}.
\end{eqnarray}
\end{widetext}
$$$$
$$$$
By using the binomial expansion and writing the hyperbolic trigonometric functions in terms of the exponential functions we get the most compact form of Eq. (\ref{eq12}) as
\begin{widetext}
\begin{eqnarray}\label{eq13}
\dot{m}_{1}&=&\frac{1}{\tau}\left(-m_{1}+\sum_{\gamma=0}^{z}\sum_{\eta=0}^{1}\Lambda_{1}(\gamma,\eta)m_{1}^{\gamma}m_{2}^{\eta}\right),\nonumber\\
\dot{m}_{2}&=&\frac{1}{\tau}\left(-m_{2}+\sum_{\gamma=0}^{z}\sum_{\eta=0}^{1}\sum_{\nu=0}^{1}\Lambda_{2}(\gamma,\eta,\nu)m_{2}^{\gamma}m_{1}^{\eta}m_{3}^{\nu}\right),\nonumber\\
\vdots\quad&\vdots&\quad\quad\quad\quad\quad\quad\quad\quad\quad\quad\vdots\nonumber\\
\dot{m}_{k}&=&\frac{1}{\tau}\left(-m_{k}+\sum_{\gamma=0}^{z}\sum_{\eta=0}^{1}\sum_{\nu=0}^{1}\Lambda_{3}(\gamma,\eta,\nu)m_{k}^{\gamma}m_{k-1}^{\eta}m_{k+1}^{\nu}\right),\nonumber\\
\vdots\quad&\vdots&\quad\quad\quad\quad\quad\quad\quad\quad\quad\quad\vdots\nonumber\\
\dot{m}_{L-1}&=&\frac{1}{\tau}\left(-m_{L-1}+\sum_{\gamma=0}^{z}\sum_{\eta=0}^{1}\sum_{\nu=0}^{1}\Lambda_{2}(\gamma,\eta,\nu)m_{L}^{\gamma}m_{L-1}^{\eta}m_{L+1}^{\nu}\right),\nonumber\\
\dot{m}_{L}&=&\frac{1}{\tau}\left(-m_{L}+\sum_{\gamma=0}^{z}\sum_{\eta=0}^{1}\Lambda_{1}(\gamma,\eta)m_{L}^{\gamma}m_{L-1}^{\eta}\right).
\end{eqnarray}
\end{widetext}
where
\begin{eqnarray}\label{eq14}
\Lambda_{1}(\gamma,\eta)&=&{{z}\choose{\gamma}}{{1}\choose{\eta}}A_{1}^{z-\gamma}A_{2}^{1-\eta}B_{1}^{\gamma}B_{2}^{\eta},\nonumber\\
\Lambda_{2}(\gamma,\eta,\nu)&=&{{z}\choose{\gamma}}{{1}\choose{\eta}}{{1}\choose{\nu}}A_{2}^{1-\eta}A_{3}^{z+1-\gamma-\nu}B_{2}^{\eta}B_{3}^{\gamma+\nu},\nonumber\\
\Lambda_{3}(\gamma,\eta,\nu)&=&{{z}\choose{\gamma}}{{1}\choose{\eta}}{{1}\choose{\nu}}A_{3}^{z+2-\gamma-\eta-\nu}B_{3}^{\gamma+\eta+\nu}.
\end{eqnarray}
Note that, throughout our calculations $\tau=1$ for simplicity. Self-consistent equations in Eq. (\ref{eq13}) are typical first order ODE but has not an analytical solution because of the right-hand side contains transcendental functions. The dynamical equation of motion can be solved by various numerical methods. In this work, we prefer to use the fourth order Runge-Kutta method (RK4) to get the evolution of the $m(t)$ by regarding Eq. (\ref{eq13}) as an initial value problem. We can mention that the differential equation derived in Eq. (\ref{eq13}) extends up to the term $m^{z}$. Each term in the equation of motion makes contribution to the solution, because the value of $m$, which is calculated at each time step, is iteratively related to the previous $m$ value, however the situation is different from the behavior of the equilibrium systems at which high ordered terms can be neglected in the neighborhood of phase transition point.

The system has three dependent Hamiltonian variables, namely frequency of external magnetic field $\omega$, amplitude $h_{0}$ and the film thickness $L$. For certain values of these parameters, temperature and the $J_{1},$ $J_{2},$ $J_{3}$ interaction constants, RK4 will give convergency behavior after some iterations i.e. the solutions have property $m(t)=m(t+2\pi/\omega)$ for arbitrary initial value for the magnetization ($m(t=0)$). Each iteration, i.e. the calculation magnetization for $t+1$ from the previous magnetization for $t$, is now performed for these purpose whereby the RK4 iterative equation is being utilized to determine the magnetization for every $i$. In order to keep the iteration procedure stable in our simulations, we have chosen $10^{4}$ points for each RK4 step. Thus, after obtaining the convergent region and some transient steps (which depends on Hamiltonian parameters and the temperature) the layer's average magnetization can be calculated from
\begin{equation}\label{eq15}
Q_{k}=\frac{\omega}{2\pi}\oint m_{k}(t)dt
\end{equation}
where $m_{k}$ is a stable and periodic function anymore and finally the DOP can be calculated by the arithmetic mean as
\begin{equation}\label{eq16}
Q=\frac{1}{L}\sum_{k=1}^{L}Q_{k}.
\end{equation}
As cut-off condition for numerical self-consistency, we defined a tolerance
\begin{equation}\label{eq17}
\left |Q_{k}|_{t-2\pi/\omega}^{t}-Q_{k}|_{t}^{t+2\pi/\omega}\right |<10^{-5},
\end{equation}
meaning that the maximum error as difference between the each consecutive iteration should be lower than $10^{-5}$ for all step. On the other hand, the modified surface exchange interaction has been defined to determine the different characteristic behavior of the system in certain range as
\begin{equation}\label{eq18}
J_{1}=J_{3}(1+\Delta_{s}).
\end{equation}

There are three possible states for the system, namely F, P and the coexistence phase (F+P). The total magnetization time-series  $m(t)$ in convergent region is satisfied by this condition
\begin{equation}\label{eq19}
m(t)=-m(t+\pi/\omega)
\end{equation}
in the P phase which is called the symmetric solution. The solution corresponding to P phase follows the external magnetic field and oscillates around zero value which means that the DOP is zero. In the F phase, the solution does not satisfy Eq. (\ref{eq19}) and this solution is called as non-symmetric solution which oscillates around a non-zero magnetization value, and does not follow the external magnetic field i.e. the value of $Q$ is different from zero. In these two cases, the observed behavior of magnetization is regardless of the choice of initial value of magnetization $m(0)$ whereas the last phase has magnetization solutions symmetric or non-symmetric depending on the choice of the initial value of magnetization corresponding to the coexistence region where F and P phases overlap. The main aim of the treatment is manifesting the frequency dispersion of the critical temperature coordinates of special point can be calculated by benefiting from the phase diagrams in $(k_{B}T/J_{3}-\Delta_{s})$ planes for different field amplitudes in order to understand and clarify the behavior of the dynamical system.

\section{Results and Discussion}\label{result}
\begin{figure}
\centering
\includegraphics[width=8.5cm]{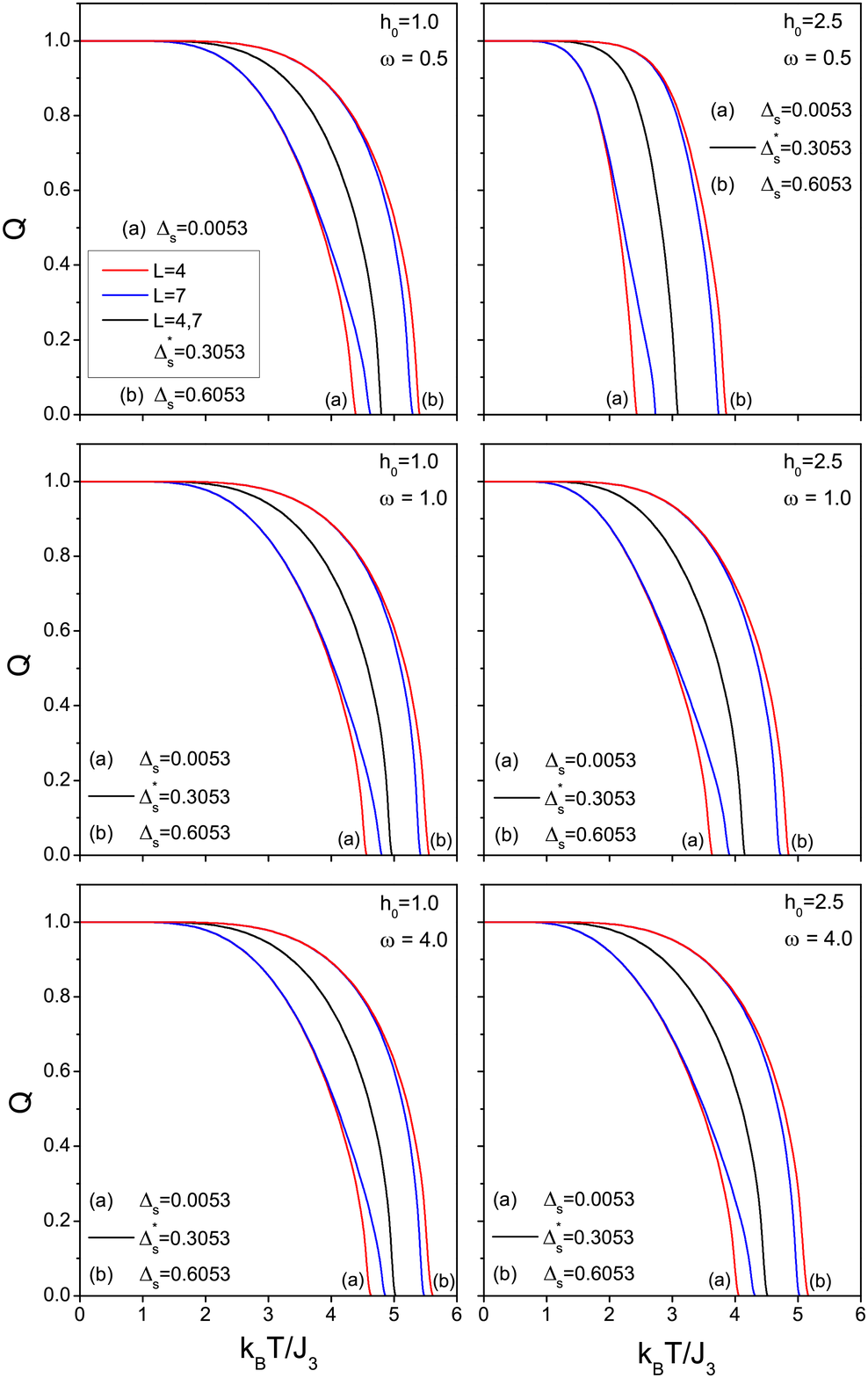}\\
\caption{(Color online) Variation of dynamical order parameter with temperature for the selected films with different $L=4,$ $7$ layers for two values of external field amplitude. The left-hand- and right-hand-triple-panels correspond to $h_{0}/J_{3}=1.0$ and $2.5$ respectively can be seen for the principal three frequency agents as $\omega=0.5,$ $1.0$ and $4.0$ stated in the text. The number accompanying each set of curves denote the values of modified exchange from (a) to (b) as $\Delta_{s}=0.0053,$ $0.3053,$ $0.6053$.}\label{fig2}
\end{figure}

\begin{figure}
\centering
\includegraphics[width=8.55cm]{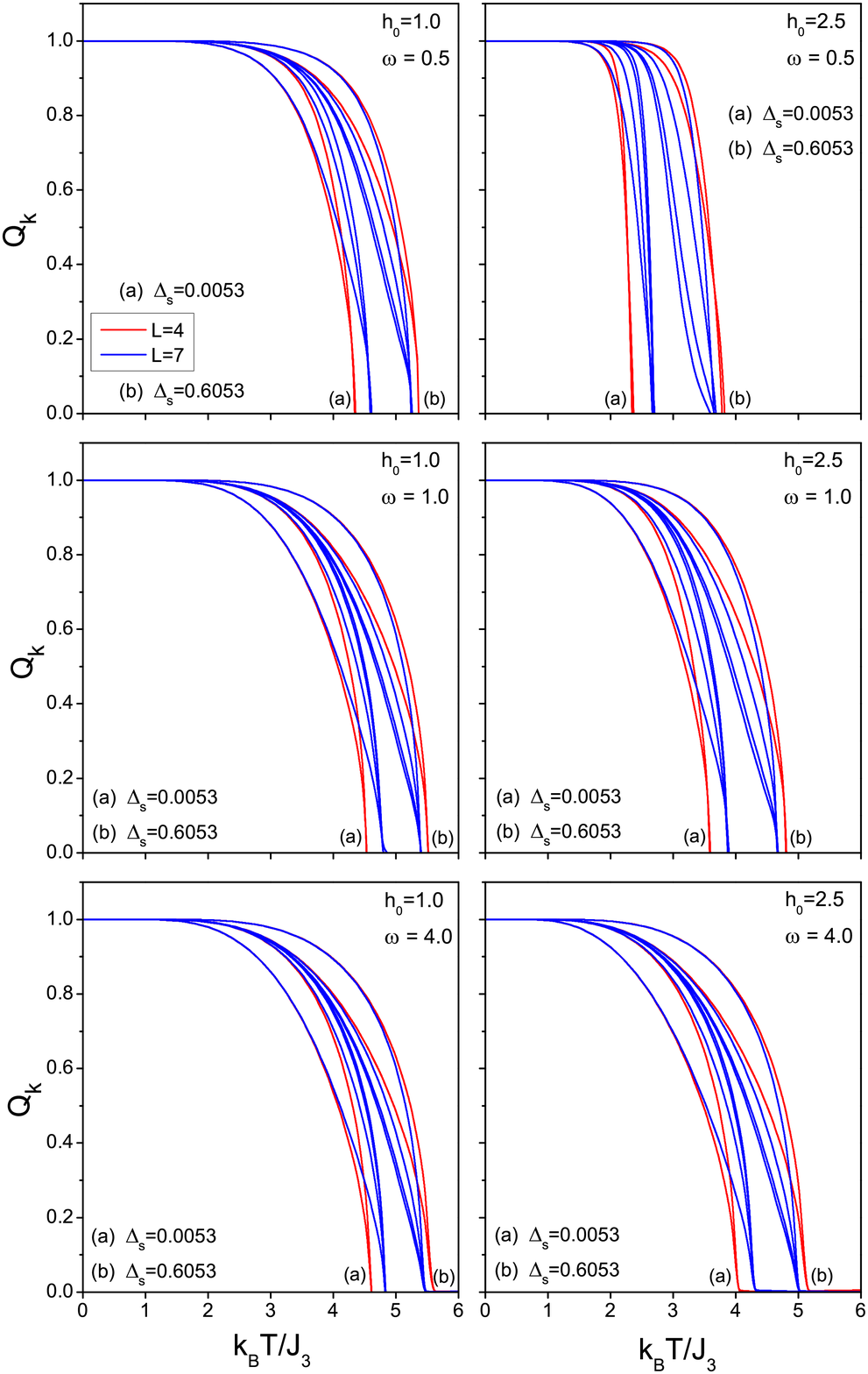}\\
\caption{(Color online) Variation of dynamical order parameter for each layers with temperature for the selected films with different $L=4,$ $7$ layers for two values of external field amplitude. The interchange of the layer hierarchy can be monitored in the figure for the same values depicted in Fig (\ref{fig2}).}\label{fig3}
\end{figure}

\begin{figure}
\centering
\includegraphics[width=8.7cm]{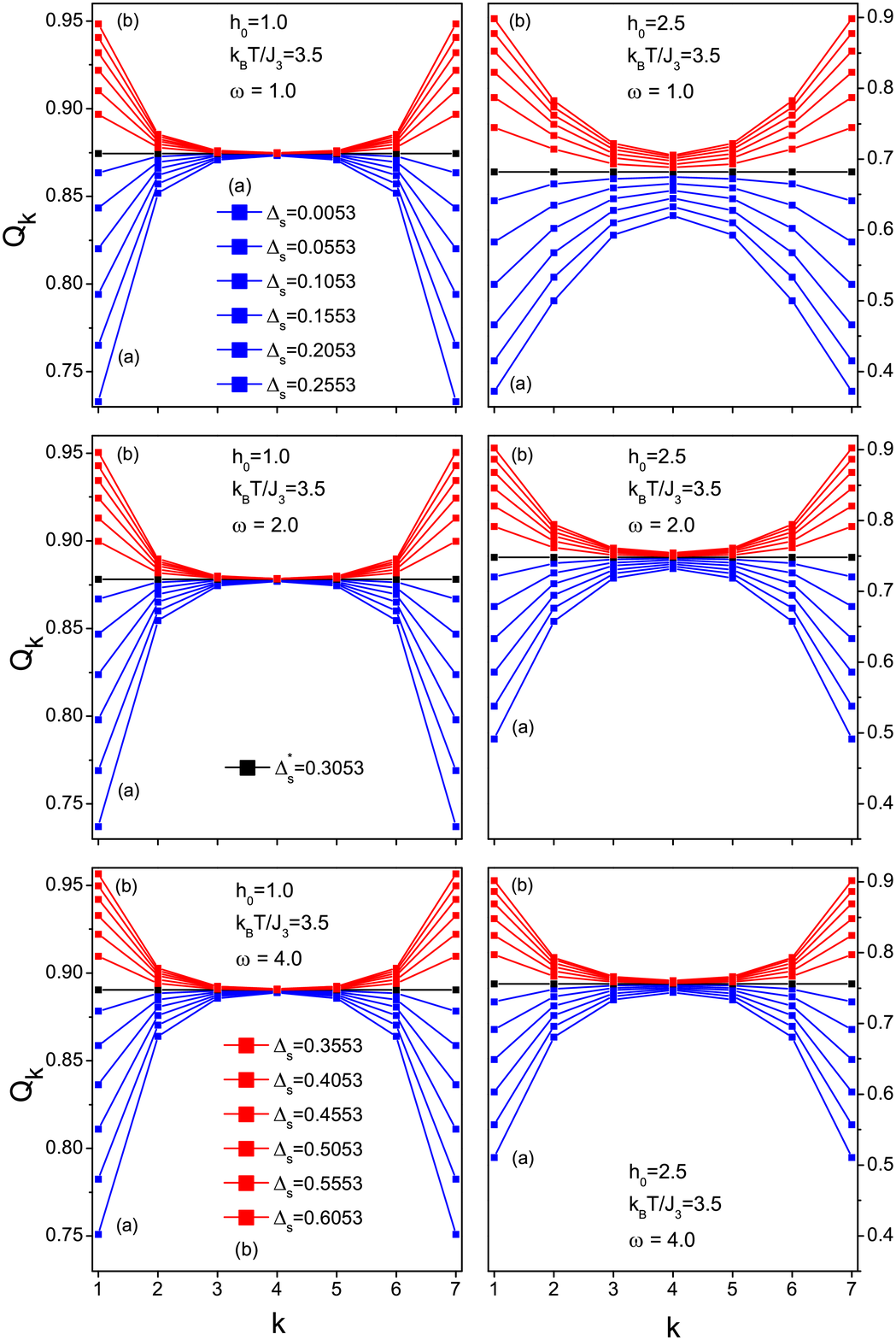}\\
\caption{(Color online) The average magnetization profiles for a film $L=7$ layers for two values of external field amplitude $h_{0}/J_{3}=1.0,$ $2.5$ and fixed temperature $k_{B}T/J_{3}=3.5$. The number accompanying each curve denotes several values of modified exchange from (a) $\Delta_{s}=0.0053$ to (b) $\Delta_{s}=0.6053$. The evolution of the curves can be seen in the left-hand- and right-hand-triple-panels for the values of frequency $\omega=0.5,$ $1.0$ and $4.0$.}\label{fig4}
\end{figure}

The best appropriate values and/or intervals of parameters are chosen in our calculations to explain the whole behavior with least-effort. Surface exchange $J_{1}$ has directly modified on $\Delta_{s}$ by passing the distinction between bulk and interface exchange couplings as $J_{2}\equiv J_{3}=1.0$. Unlike the deductive analysis, presenting the inferences as a result of investigation by induction is an essential principle in terms of understanding the overall behavior for such a scrutiny. In this sense, variation of DOP with temperature for the selected different film thicknesses $L=4,$ $7$ which has been rigorously depicted in the text, constitutes as the starting point for our systematic investigation. The preliminary results are presented in Fig. (\ref{fig2}) for two selected values of external field amplitude and for the principal three frequency agents such as $h_{0}/J_{3}=1.0,$ $2.5$ and $\omega=0.5,$ $1.0,$ $4.0$ respectively. It is worth noting that the value of the crossover point $\Delta_{s}^{*}$ has been calculated as $0.3053$ within the EFT in accordance with value reported by Kaneyoshi \cite{kaneyoshi4}. A little bit the left and the right of this point at $\Delta_{s}=0.0053,$ and $0.6053$ respectively, corresponding average layer magnetizations have contra-arranged strength of the magnetic order mutually. Thus, the three representatives value of the modified exchange also have been taken and denoted from (a) to (b) as competence parameters $\Delta_{s}=0.0053,$ $0.3053,$ and $0.6053$. We also point out that this labeling procedure can be used for usual relevant behaviors in the same investigation from a different viewpoint, i.e. it has been used for temperature dependency of the average magnetization of each layer in Fig. (\ref{fig3}) and the variation of it with the layer index $k$ at a fixed temperature where $k_{B}T/J_{3}=3.5$ in Fig. (\ref{fig4}). Average magnetization profiles across the film differ qualitatively in two regimes ($\Delta_{s}< \Delta_{s}^{*}$ and $\Delta_{s}> \Delta_{s}^{*}$) as shown in Fig. (\ref{fig4}) for the amplitude agents. Apart from this, DOP and consequently the critical temperature is independent from thickness $L$ in the crossover point and this makes it `special'. The thickness-independent critical temperature of the film at special point reduces to the bulk with coordination number $z=6$ critical value $k_{B}T/J_{3}=5.0734$ in infinite-frequency for particular amplitude value or zero-amplitude limit while there is no significant change in the value of the $\Delta_{s}^{*}$ when $w$ or $h_{0}$ chances. This reduction is one of the major indicators for the accuracy of our treatment. The effect of external oscillatory magnetic field (frequency and amplitude of dynamic field) on critical behavior of the bulk systems is now well known. At first sight, we can see in Fig. (\ref{fig2}), (\ref{fig3}), and Fig. (\ref{fig4}) that a weak increase in $\omega$ causes an increasing in $k_{B}T_{c}/J_{3}$. However, an increasing in $h_{0}/J_{3}$ causes a decreasing $k_{B}T_{c}/J_{3}$ for given $\Delta_{s}$ and arbitrary $\omega$. This is an expected result, since the increment in frequencies enhances the phase lag between time dependent magnetization and the external field signal. Therefore, the asymmetric behavior of HL in the form of Lissajous curve, becomes more prominent, since the time dependent magnetization has less time to follow the oscillatory field. So, the system can undergoes a DPT which requires a small amount of thermal energy. On the other hand, according to times-series of the magnetization and the external magnetic field, increasing the field frequency at first, obstructs the saturation of the ordinary magnetization due to the decreasing energy coming from the oscillating magnetic field in a half-time period which facilitates the late stage domain growth by tending to align the moments in its direction (i.e. the magnetization begins to fail following the oscillatory field) and this makes the occurrence of the frequency increasing route to DPT at the critical point. There is a concurrence by different researchers that the HL loses its symmetry when the oscillating period of external perturbation becomes much smaller than the typical relaxation time of the system.
\begin{figure}[b]
\centering
\includegraphics[width=8.5cm]{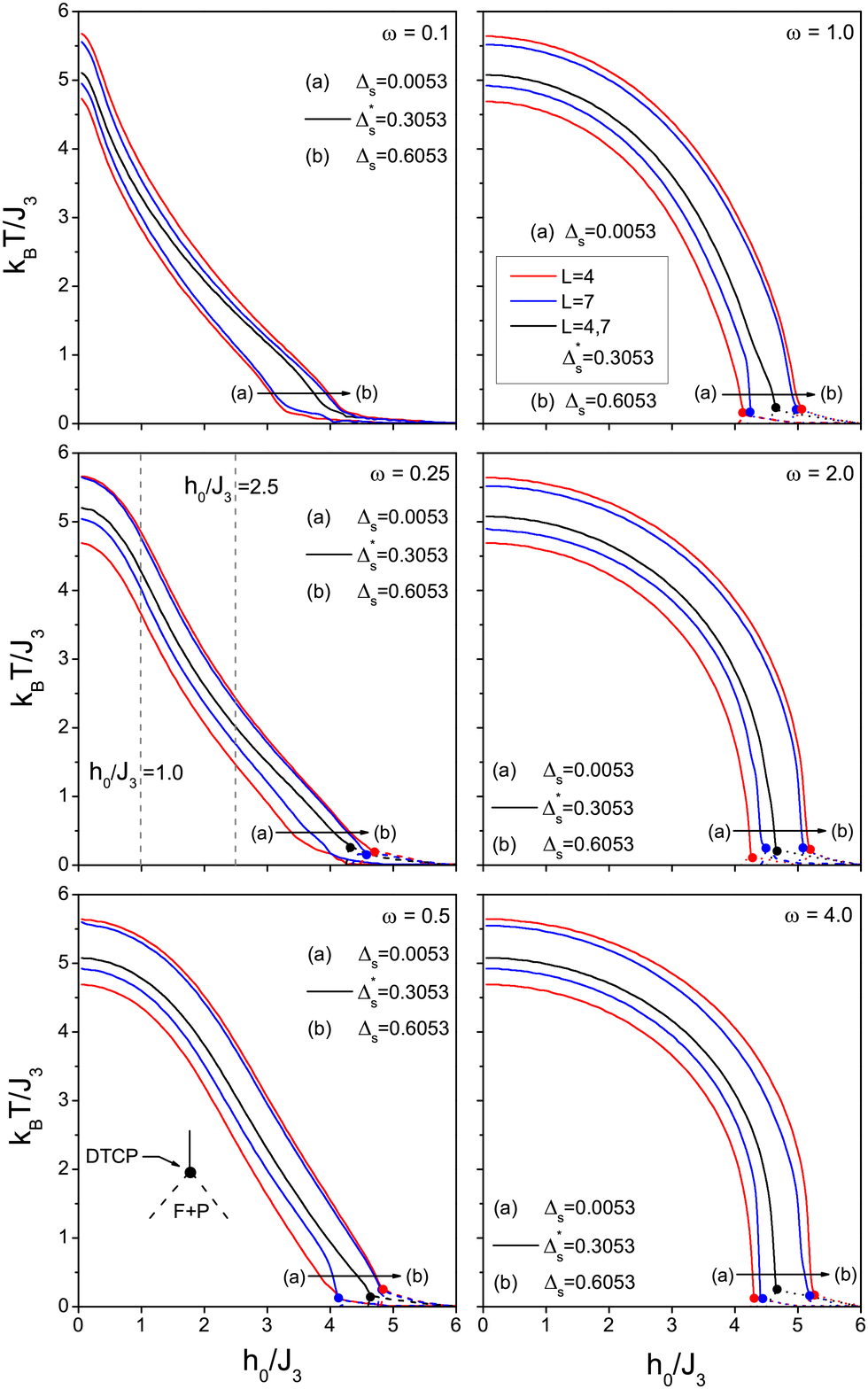}\\
\caption{(Color online) The phase diagram in the $(h_{0}/J_{3}-k_{B}T/J_{3})$ plane for the films with different $L=4,$ $7$ layers. Frequency changes with $\omega=0.1, 0.25, 0.5, 1.0, 2.0$ and $4.0$ for each panels. The numbers accompanying each set of curves denote the value of modified exchange from (a) to (b) as $\Delta_{s}=0.0053,$ $0.3053,$ $0.6053$ and this type of display format is also stated as an arrow in the figure. The dashed lines in the middle-left-hand panel corresponds to the representatives serve as a viewing guide.}\label{fig5}
\end{figure}

The above-mentioned mechanism is due to the competing time scales in such non-equilibrium systems and also can be seen in the global phase diagrams in ($h_{0}/J_{3}-k_{B}T/J_{3}$) planes which are presented in Fig. (\ref{fig5}). The external oscillatory field defined in Eq. (\ref{eq2}) is governed by only an open variable $h_{0}$ which is the amplitude of the field. However, the system has close dependency to frequency as embedded into a transcendental function. Although the time average of the field over a full period is zero, sinusoidally oscillating field drags the system to the ordered phase. In addition to this, bulk and modified exchange interactions enforce the system to stay in the ordered phase. The last factor, the temperature, which causes thermal agitations, induces a disordered phase when the energy supplied by the temperature to the system is high enough. Thus, mainly a competition takes place between the amplitude and temperature. After this brief summary about the physical background of critical phenomena observed in the dynamical system, let us investigate the effects of external field frequency and modified exchange on the selected different films thicknesses $L=4,$ $7$. In Fig. (\ref{fig5}), phase diagrams can be seen for the relevant thickness $L$ with chosen $\Delta_{s}$. At first glance, one can see that the Ising-type pure crystalline thin film exhibits meta-stable behavior. A storable coexistence region shows itself at low temperature and high amplitude values where the F and P phases overlap and the critical properties of the system depends on the initial value of the magnetization. If the frequency value increases such meta-stable phases appear. Contrary to this, the coexistence (F+P) regions disappear with decreasing frequency at the high temperature and low amplitude values. Although there is a noteworthy information about the meta-stability, probably it would be wrong to skip the unpretentious remark that the meta-stable regions also have the maximal area in the crossover value of modified exchange in the diagrams. The representative amplitudes has already been selected as $h_{0}/J_{3}=1.0$ and $2.5$ privately to avoid complications of coexistence regions that may be an artifact phenomenon of the effective-field approach itself reported several times by some researchers \cite{aktas, idigoras, berkolaiko, shi}. In addition, there exists a dynamical tricritical point on the dynamical phase boundary at which the second order and first order phase transition lines are separated from each other. Besides, rising frequency leads to the flaring curves in the phase planes, contrary to decreasing frequency shrinks the ferromagnetic region. This reflects the truth, since increasing the field frequency causes a growing phase delay between the magnetization and field, then this makes the occurrence of the DPT difficult. As a result of this mechanism the dynamical phase boundary (DPB) gets wider. The modified exchange has been chosen as the best appropriate values $\Delta_{s}=0.0053,$ $0.3053,$ $0.6053$ with reference to above procedure (which provide the conditions $\Delta_{s}<\Delta_{s}^{*}$ and $\Delta_{s}^{*}<\Delta_{s}$, respectively). It is clear that the results presented in Fig. (\ref{fig5}) are completely consistent with the results shown in Fig. (\ref{fig2}), (\ref{fig3}), and Fig. (\ref{fig4}) since the $\Delta_{s}$ comes close to the crossover value then the behavior of the system begins to resemble as quasistatic (the phrase is coming from the fact that the HLs has a nonzero residual loop area, originates from the deficiency of the relevant method, even in the zero-frequency limit). Also, the interchanging behavior can be seen from the critical value of the transition in inner- and outer-stablemate-curves around the characteristic one corresponding crossover. Physical mechanism mentioned above can be briefly explained as follows: If one keeps the system in one well of a Landau type double well potential, a certain amount of energy originating from magnetic field is necessary to achieve a dynamic symmetry breaking. If the amplitude of the applied field is less than the required amount then the system oscillates in one well. In this situation, the magnetization does not change its sign. In other words, the system oscillates around a nonzero value. This region is dynamically ordered phase. When the temperature increases, the height of the barrier between the two wells decreases. As a result of this, the less amount of magnetic field is necessary to push the system from one well to another and hence the magnetization can change its sign for this amount of field. Consequently, the time averaged magnetization over a full cycle of the oscillating field becomes zero. In addition, we can mention that for a dynamical thin film, increasing the dipole-dipole interaction-induced energy contribution by rising the strength of the modified surface exchange interaction will cause an increase the orderliness tendency of the system which makes the transition from one well to another more difficult.
\begin{figure}
\centering
\includegraphics[width=8.6cm]{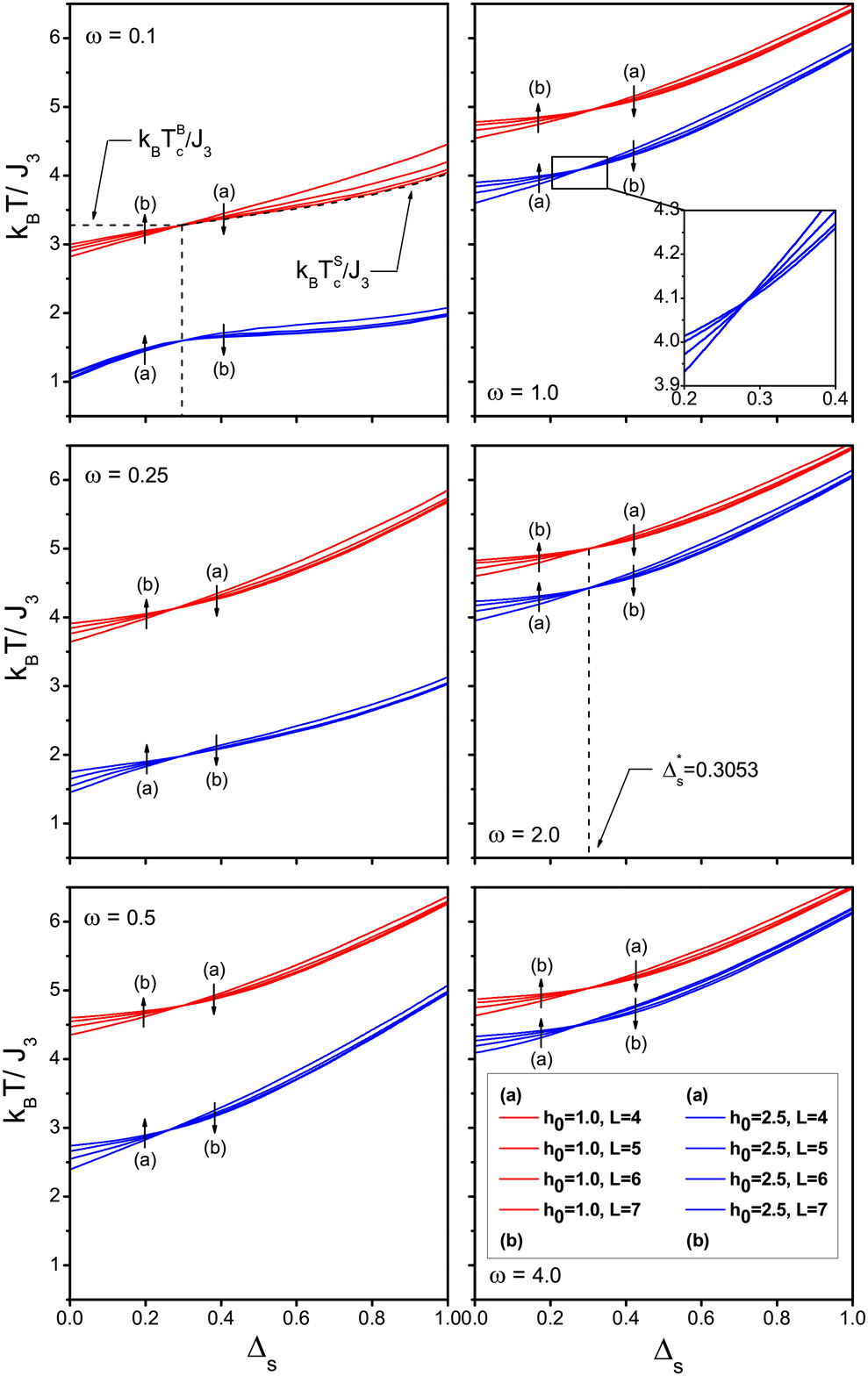}\\
\caption{(Color online) The critical temperature versus modified exchange in $(\Delta_{s}-k_{B}T/J_{3})$ phase planes for the films with different $L=4,$ $5,$ $6,$ and $7$ layers. Two set of curves correspond to the selected amplitude agents $h_{0}/J_{3}=1.0$ and $2.5$. The number accompanying each panels denote the value of frequency $\omega=0.1, 0.25, 0.5, 1.0, 2.0$ and $4.0$. Differences in the layer hierarchy on both sides of the crossover point is shown by the arrows in opposite directions. The exemplary inset of the crossover point is stated in terms of visual clarity.}\label{fig6}
\end{figure}

As stated in various place, a prime objective of the study was to confirm whether a common value of $\Delta_{s}$ existed for all films, and if so to determine its value. This requires a very accurate determination of the transition temperature for each film thickness and each value of surface exchange enhancement. For this purpose, we need to dart a glance at the relevant ($\Delta_{s}-k_{B}T/J_{3}$) global phase diagrams are shown in Fig. (\ref{fig6}). They are plotted for different numbers of layers when the strength of the field amplitude is $h_{0}/J_{3}=1.0$, and $2.5$ respectively. As $\Delta_{s}$ rises, the critical temperature of the system increases for arbitrary $L$. For each value of the frequency with the representative amplitudes, as seen clearly that the curves with different film thicknesses intersect each other at the same abscissa point $\Delta_{s}=\Delta_{s}^{*}=0.3053$. According to this result, the special point can be defined as that particular $\Delta_{s}$ value at which bulk critical temperature is independent of the film thickness $L$ occurs. It is assumed to coincide with the critical temperature $k_{B}T_{c}^{B}/J_{3}$ of the corresponding isometric lattice (usual simple cubic lattice for the system under consideration). Furthermore, based on the definition of $\Delta_{s}$, it can be expected that the crossover point in Fig. (\ref{fig6}) should define also the critical temperature of three-dimensional infinite bulk system, where the surface and the $\Delta_{s}$ parameter are of no importance. This is really the case, which can be seen also from Fig. (\ref{fig6}), where the bulk $k_{B}T_{c}^{B}/J_{3}$ and the surface $k_{B}T_{c}^{S}/J_{3}$ critical temperatures of the corresponding semi-infinite system are represented respectively by the dashed lines and also pointed by one each arrow. For the values of $\Delta_{s}<\Delta_{s}^{*}$, thin films have lower critical temperature than the bulk system. The value of $\Delta_{s}$ also changes the relation between the thickness and the critical temperature of the film. $k_{B}T_{c}/J_{3}$ increases with the film thickness $L$, and approaches $k_{B}T_{c}^{B}/J_{3}$ asymptotically as the number of layers become large. The thicker films have higher critical values for $\Delta_{s}<\Delta_{s}^{*}$ whereas they have lower critical values for $\Delta_{s}>\Delta_{s}^{*}$ than thinner ones. It is worth noting once again that the critical temperature of the film is independent of $L$ when $\Delta_{s}=\Delta_{s}^{*}$, and equal to $k_{B}T_{c}^{B}/J_{3}$. On the other hand, for $\Delta_{s}>\Delta_{s}^{*}$ the critical temperature $k_{B}T_{c}/J_{3}$ is greater than both the bulk $k_{B}T_{c}^{B}/J_{3}$ and the surface $k_{B}T_{c}^{S}/J_{3}$ critical temperatures of the corresponding semi-infinite Ising system and larger the $L$ is, the lower $k_{B}T_{c}/J_{3}$ is. The film critical temperature $k_{B}T_{c}/J_{3}$  approaches asymptotically the surface critical temperature $k_{B}T_{c}^{S}/J_{3}$ of the corresponding semi-infinite system as the number of layers become large. Beyond all results, the evolution of the curves with respect to frequency as $\omega=0.1,$ $0.25,$ $0.5,$ $1.0,$ $2.0,$ and $4.0$ for particular two amplitude agents as stated several times above $h_{0}/J_{3}=1.0,$ and $2.5$ can observed in Fig. (\ref{fig6}). Accordingly, causes all the curves go-up and -down with rising $\omega$/$h_{0}/J_{3}$ respectively also the same operation increases/decreases the critical temperature of the system while it does no affect the relation between the film thickness and critical temperature, i.e. for $\Delta_{s}<\Delta_{s}^{*}$ thicker films have higher critical values and reverse is valid for $\Delta_{s}>\Delta_{s}^{*}$ with $\omega,$ $h_{0} \neq 0$, if the special point present. In addition, there is no significant change in the abscissa point of the crossover ($\Delta_{s}^{*}$) while rising or lowering of two oscillatory parameter transports the phase diagrams collectively in the space. When $\omega$ is large enough for finite amplitude value as $h_{0} \neq 0$, the ordinate value of the special point approaches the bulk critical temperature in quasistatic case $k_{B}T_{c}^{B}/J_{3}$ asymptotically. The observations about dynamical feature of the global phase diagrams in ($\Delta_{s}-k_{B}T/J_{3}$) planes can also be explained by the well known mechanism underlying the dynamical phase transitions phenomena. In other words, the aforementioned scenario is onset as things stand.

\begin{figure}
\centering
\includegraphics[width=8cm]{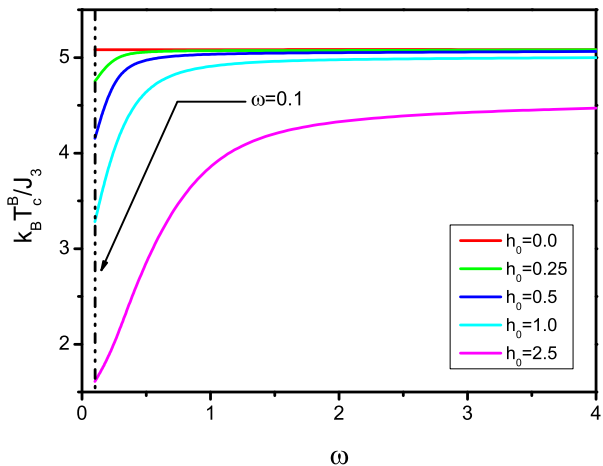}\\
\caption{(Color online) The frequency dispersion of the critical temperature coordinate of the special point for different values of the amplitude of external oscillatory field $h_{0}/J_{3}=0.0,$ $0.25,$ $0.5,$ $1.0$ and $2.5$. The dashed line at $\omega=0.1$ serves as a boundary for quasistatic limit.}\label{fig7}
\end{figure}
As one of the most striking aspects of our work is that the frequency dispersions of the critical temperature coordinates of special point for various external oscillatory field amplitudes as $h_{0}/J_{3}=0.0,$ $0.25,$ $0.5,$ $1.0,$ $2.5$ shown in Fig. (\ref{fig7}). In accordance with our anticipations, the dispersion curves evolve out of the characteristic line which accounts for the pure bulk system in quasitatic limit as rising $h_{0}/J_{3}$. The behavior of the dispersions qualitatively supports the last parts of the descriptions in the previous paragraph that is to say all dispersion curves saturates to the value of bulk critical temperature in $\omega \rightarrow \infty$ limit. Contrary to this, they go to a finite non-zero value in $\omega \rightarrow 0$ limit. Major attention has been paid to catch the overall behavior of a dynamical thin film in our calculations, hence the lowest value of the frequency in the treatment is $\omega=0.1$. This value is sufficient for estimation an existence of the HL residuality mentioned above even now.

\section{Conclusion}\label{conclude}
In this study, we have investigated the surface enhancement phenomenon and dynamic nature of the critical phenomena which is observed for dynamical Ising-type thin films with various thickness which has been defined with inner coordination number $z=4$ driven by an external oscillatory magnetic field by means of effective-field theory based on a standard decoupling approximation. The time evolution of the system has been presented by utilizing a Glauber type stochastic process. Our starting point for the systematic review was to examine the DOP, average magnetization of layers with respect to temperature and layer index for the selected field amplitude and frequency values at finite temperature below the Curie point of the static case on a pure Ising-type bulk. In addition to the observations set forth in the literature, the effect of the frequency and amplitude on the standard arguments have been propounded. For this purpose, the best appropriate values and/or intervals of parameters which are given in the text, have been chosen for the treatment. We supported the relevant investigations with phase diagrams globally as well as the average magnetization.

In order to examine the effect of the dynamical parameters of external field on the phase diagrams, firstly we presented the phase diagrams for selected three distinct modified exchange representatives with three values of frequency in ($h_{0}/J_{3}-k_{B}T/J_{3}$) planes. In the immediate aftermath, the frequency induced evolution of the corresponding agents along the routes at $h_{0}/J_{3}=1.0,$ $2.5$ respectively in the previous phase diagrams have been investigated at finite temperature in ($\Delta_{s}-k_{B}T/J_{3}$). According to our findings, a weak decreasing of the amplitude causes decreasing in $\Delta_{s}$ coordinate of the crossover point, while a weak increasing of the frequency causes an increasing critical temperature component of the crossover. For a better view, we have depicted frequency dispersion of the $k_{B}T_{c}/J_{3}$ coordinate of the special point for aforementioned amplitude agents.

EFT takes the standard MFA predictions one step forward by taking into account the single spin correlations which means that the thermal fluctuations are partially considered. Although all of the observations reported in this work shows that EFT can be successfully applied to such nonequilibrium realistic systems, the true nature of the physical facts underlying the observations displayed in the system (especially the origin of the different contrary effects of two variables of the external filed related between the competing time scales) may be further understood with an improved version of the present EFT formalism which can be achieved by attempting to consider the multi spin correlations which originate when expanding the spin identities or a MC simulation as stochastic process. We believe that this attempt could provide a treatment beyond the present approximation.

The present work can be helpful as first approximation on this dynamical problem. In conclusion, we hope that the results obtained in this work would shed light on the further investigations of the dynamic nature of the critical phenomena in pure crystalline systems (e.g. thin films and semi-infinite systems) and would be beneficial from both theoretical and experimental points of view.

\section*{Acknowledgements}
The numerical calculations reported in this paper were performed at T\"{U}B\.{I}TAK ULAKB\.{I}M (Turkish agency), High Performance and Grid Computing Center (TRUBA Resources) and this study has been completed at Dokuz Eyl\"{u}l University, Graduate School of Natural and Applied Sciences. One of the authors (B.O.A.) would like to thank the Turkish Educational Foundation (TEV) for full scholarship.

\end{document}